\documentstyle[preprint,prd,aps,feynmf]{revtex}
\newcommand{\be}{\begin{equation}}
\newcommand{\ee}{\end{equation}}
\newcommand{\bea}{\begin{eqnarray}}
\newcommand{\eea}{\end{eqnarray}}
\newcommand{\sptwo}{1.4}
\newcommand{\doublespace}{\edef\baselinestretch{\sptwo}\Large\normalsize}
\newcommand{\newsection}[1]{\setcounter{equation}{0}}

\newcounter{newapp}
\begin{document}
\begin{titlepage}
\vspace*{1.0in}
\begin{center}
{\bf Muonium-Antimuonium oscillations and massive Majorana neutrinos}\\
\end{center}
\begin{center} 
T.E. Clark\footnote{e-mail address: clark@physics.purdue.edu} and S.T. Love\footnote{e-mail address: love@physics.purdue.edu} \\
{\it Department of Physics\\ 
Purdue University\\
West Lafayette, IN 47907-2306}
~\\
\end{center}
\vspace{1in}
\begin{center}
{\bf Abstract}\\
The electron and muon number violating muonium-antimuonium oscillation process can proceed provided neutrinos have non-zero masses and mix among the various generations. Modifying the Standard Model only by the inclusion of singlet right handed neutrino fields and allowing for general neutrino masses and mixings, the leading order matrix element contributing to this process is computed. For the particularly interesting case where the neutrino masses are generated by a see-saw mechanism with a very large Majorana mass $M_R>>M_W$, it is found that both the very light and very heavy Majorana neutrinos each give comparable contributions to the oscillation time scale proportional to $M_R^2$. Present experimental limits set by the non-observation of the oscillation process sets a lower limit on $M_R$ of roughly of order $10^4$ GeV. 

\end{center}

\end{titlepage}

\doublespace

One of the most striking examples of a purely quantum mechanical phenomenon which occurs in a wide range of physical systems involves the time dependent oscillation between two distinct levels or particle species. Examples of such systems are quite varied and range from the text book example of a particle moving in a double well potential of the ammonia molecule to oscillations in the neutral $K^0-\overline{K}^0$ and $B^0-\overline{B}^0$ meson systems\cite{R}-\cite{no}. The later processes arise as a consequence of the fact that the left handed quarks which participate in the weak interaction are linear combinations of the mass diagonal quark states. Thus the weak interaction currents coupling to the $W^\pm$ vector bosons when written in terms of the quark mass eigenstates contain the left handed charge 2/3 quarks and a mixture of the left handed charge -1/3 quarks. Thus the neutral meson can convert into its antiparticle via the exchange of two $W$ vector bosons which arises in fourth order in the weak interaction.

Of the various fermions appearing in the Standard Model, the only ones which are electrically neutral are the neutrinos. Left-handed neutrinos are components of  $SU(2)_L$ doublets along with their charged leptonic partners and experience only the weak interaction, while any right-handed neutrinos are completely neutral under the Standard Model gauge group. During the past several years, there has been mounting experimental evidence\cite{enm1}-\cite{enm4} that the neutrinos involved in the weak interactions are in fact linear combinations of nonzero mass neutrino eigenstates in a somewhat analogous fashion to the quarks. Thus the various weak interaction neutrinos exhibit a mixing phenomenon and, being electrically neutral, can also oscillate from one species into another. The size and nature of the neutrino mass and the associated mixing is still an open question subject to experimental determination and theoretical speculation\cite{tnm1}. 

Muonium ($M$)\cite{P} is the Coulombic bound state of the electron and the antimuon ($e^-\mu^+$), while antimuonium ($\overline{M}$) is the Coulombic bound state of the positron and the muon ($e^+\mu^-$). In order for there to be a nontrivial mixing between these two states, the individual electron and muon number conservation must be violated. Such a situation results provided the neutrinos are massive particles which mix amongst the various generations. Modifying the Standard Model only by the inclusion of singlet right handed neutrinos and allowing for a general mass matrix for the neutrinos, this criterion can be met and the $e^- \mu^+$ and $e^+\mu^-$ states can indeed mix\cite{RHC}.

The lowest order Feynman diagrams producing  muonium and antimuonium mixing are displayed in Figure 1. 

\vspace*{0.4in}
\begin{fmffile}{muantimua}
\begin{fmfgraph*}(40,30)
\fmfpen{thick}
\fmfleft{i1,i2}
\fmflabel{$e$}{i1}
\fmflabel{$\mu$}{i2}
\fmfright{o1,o2}
\fmflabel{$\mu$}{o1}
\fmflabel{$e$}{o2}
\fmf{fermion}{i1,v1}
\fmf{photon,tension=.4,label=$W$}{v1,v3}
\fmf{fermion}{v3,o1}
\fmf{fermion}{o2,v4}
\fmf{photon,tension=.4,label=$W$}{v4,v2}
\fmf{fermion}{v2,i2}
\fmf{plain,tension=.4,label=$\nu_A$,label.side=left}{v1,v2}
\fmf{plain,tension=.4,label=$\nu_B$,label.side=right}{v3,v4}
\fmfdotn{v}{4}
\end{fmfgraph*}
\end{fmffile}
\begin{fmffile}{muantimub}
\begin{fmfgraph*}(40,30)
\fmfpen{thick}
\fmfleft{i1,i2}
\fmflabel{$e$}{i1}
\fmflabel{$\mu$}{i2}
\fmfright{o1,o2}
\fmflabel{$\mu$}{o1}
\fmflabel{$e$}{o2}
\fmf{fermion}{i1,v1}
\fmf{plain,tension=.4,label=$\nu_B$,label.side=right}{v1,v3}
\fmf{fermion}{v3,o1}
\fmf{fermion}{o2,v4}
\fmf{plain,tension=.4,label=$\nu_A$,label.side=right}{v4,v2}
\fmf{fermion}{v2,i2}
\fmf{photon,tension=.4,label=$W$,label.side=left}{v1,v2}
\fmf{photon,tension=.4,label=$W$,label.side=right}{v3,v4}
\fmfdotn{v}{4}
\end{fmfgraph*}
\end{fmffile}
\linebreak

\vspace{0.25in}

\noindent
\hspace{1.0in}(a)\hspace{3.7in}(b)\hfil
\linebreak

\vspace*{0.25in}
\begin{fmffile}{muantimuc}
\begin{fmfgraph*}(40,30)
\fmfpen{thick}
\fmfleft{i1,i2}
\fmflabel{$e$}{i1}
\fmflabel{$\mu$}{i2}
\fmfright{o1,o2}
\fmflabel{$e$}{o1}
\fmflabel{$\mu$}{o2}
\fmf{fermion}{i1,v1}
\fmf{plain,tension=.4,label=$\nu_B$,label.side=right}{v1,v3}
\fmf{fermion}{o1,v3}
\fmf{fermion}{v4,o2}
\fmf{plain,tension=.4,label=$\nu_A$,label.side=right}{v4,v2}
\fmf{fermion}{v2,i2}
\fmf{photon,tension=.4,label=$W$,label.side=left}{v1,v2}
\fmf{photon,tension=.4,label=$W$,label.side=right}{v3,v4}
\fmfdotn{v}{4}
\end{fmfgraph*}
\end{fmffile}
\begin{fmffile}{muantimud}
\begin{fmfgraph*}(40,30)
\fmfpen{thick}
\fmfleft{i1,i2}
\fmflabel{$e$}{i1}
\fmflabel{$\mu$}{i2}
\fmfright{o1,o2}
\fmflabel{$e$}{o1}
\fmflabel{$\mu$}{o2}
\fmf{fermion}{i1,v1}
\fmf{plain,tension=.3,label=$\nu_B$,label.side=right}{v1,v3}
\fmf{fermion}{o1,v3}
\fmf{fermion}{v4,o2}
\fmf{plain,tension=.3,label=$\nu_A$,label.side=right}{v4,v2}
\fmf{fermion}{v2,i2}
\fmf{photon,tension=.4}{v1,v4}
\fmf{photon,tension=.4,rubout}{v2,v3}
\fmfdotn{v}{4}
\fmfv{label=$W$,l.a=-90,l.d=.1w}{v2}
\fmfv{label=$W$,l.a=-90,l.d=.1w}{v4}
\end{fmfgraph*}
\end{fmffile}
\linebreak

\vspace{0.25in}

\noindent
\hspace{1.0in}(c)\hspace{3.7in}(d)\hfil
\linebreak

\begin{center}
\noindent
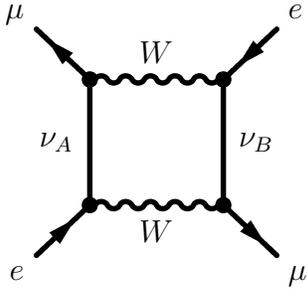
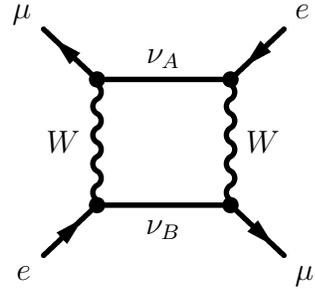
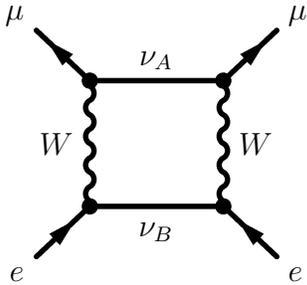
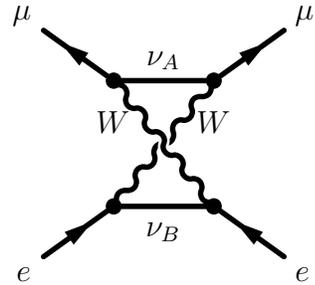
{\bf Figure 1:} {Feynman graphs contributing to the muonium-antimuonium
mixing.}
\end{center}

\noindent
They contain two neutrinos and two $W$ bosons in the intermediate state. The first two graphs ($a$ and $b$) contributing to the muonium-antimuonium mixing bears a striking resemblance to the quark box diagram underlying the mixing in the neutral kaon and B meson systems\cite{tm1}. For the  $M-\overline{M}$ case, however, one is free from the complications of the strong interactions since both $M$ and $\overline{M}$ are simple nonrelativistic Coulombic bound states composed of leptonic constituents. Thus the mixing of the $M-\overline{M}$ system can be unambiguously calculated. On the other hand, the later two graphical contributions ($c$ and $d$) arise since, in general, the neutrino mass eigenstates are Majorana (self-conjugate). These graphs have no counterpart in the quark case. As it turns out, however, these graphs cancel against each other in the calculation of the effective Lagrangian and thus do not contribute to the muonium-antimuonium oscillation process.  

The Standard Model leptonic charged current interaction is given by 
\be
{\cal L}_{cc}^{lept} =\frac{e}{sin \theta_W} (J_{-\mu }^{lept}W^{+ \mu} + J_{+\mu }^{lept}W^{- \mu}) ~,
\ee
where the (purely left-handed) leptonic charged current is 
\be
J_{-\mu }^{lept}=(J_{+\mu }^{lept})^\dagger =\sum_{a=1}^3 \overline{\ell_{La}^{(0)}}\gamma_\mu \nu_{La}^{(0)} ~.
\ee
Here $\ell_{La}^{(0)}$ and $\nu_{La}^{(0)}$ are respectively the charged lepton and its associated neutrino partner of the $SU(2)_L$ doublet. The superscript zero indicates  weak interaction eigenstates. A generic left handed fermion field is $\psi_L=\frac{1}{2}(1-\gamma_5)\psi $, while  the right handed fermion field is $\psi_R=\frac{1}{2}(1+\gamma_5)\psi $. 
The fields participating in the weak interaction, however, need not be mass diagonal. The mass term for the charged leptons arises from their Yukawa couplings with the scalar doublet. After spontaneous symmetry breaking, the mass term takes the form 
\be
{\cal L}^{\ell}_{mass} = -\sum_{a,b=1}^3 [\overline{\ell_{Ra}^{(0)}}m_{ab}^{\ell}\ell^{(0)}_{Lb}+
\overline{\ell_{La}^{(0)}}m_{ba}^{\ell *}\ell^{(0)}_{Rb}] ~,
\ee
where ${\bf m^\ell}$ is a $3\times 3$ mass matrix. To diagonalize this matrix, one performs a biunitary transformation
\be
{\bf m^\ell} ={\bf A^R} {\bf m^\ell_{diag}}{\bf (A^L)}^\dagger ~,
\ee
where ${\bf A^R}$ and ${\bf A^L}$ are $3\times 3$ unitary matrices and ${\bf m^\ell_{diag}}$ is a diagonal $3\times 3$ matrix whose entries are the charged lepton masses, $m_{\ell a}$. To implement this basis change, the charged lepton fields participating in the weak interaction are rewritten in terms of the mass diagonal fields (denoted without the superscript) as 
\be
\ell^{(0)}_{La}= \sum_{a=1}^3 A^{L}_{ab}\ell_{Lb}~~;~~\ell^{(0)}_{Ra}= \sum_{a=1}^3 A^{R}_{ab}\ell_{Rb} ~.
\ee
So doing the mass term reads
\be
{\cal L}^{\ell}_{mass} = -\sum_{a=1}^3 m_{\ell a}[\overline{\ell}_{Ra}\ell_{La}+\overline{\ell}_{La}\ell_{Ra}] ~.
\ee

A general neutrino mass term takes the form
\bea
{\cal L}_{\nu}^{mass}&=&-\frac{1}{2}\sum_{a,b=1}^3 [\nu^{(0)T}_{La}Cm^{L}_{ab}\nu^{(0)}_{Lb}+ \overline{\nu_{La}^{(0)T}}Cm^{L*}_{ba}\overline{\nu_{Lb}^{(0)}}]-\sum_{a=1}^3\sum_{i=4}^6 [\overline{\nu_{Ri}^{(0)}}m^{D}_{ia}\nu_{La}^{(0)} + \overline{\nu_{La}^{(0)}}m^{D*}_{ia}\nu^{(0)}_{Ri}]\nonumber\\
&&-\frac{1}{2}\sum_{i,j=4}^6 [\overline{\nu_{Ri}^{(0)}}Cm^{R}_{ij}\overline{\nu_{Rj}^{(0)T}}+\nu_{Ri}^{(0)T}Cm^{R*}_{ji}\nu^{(0)}_{Rj}] ~.
\eea
Here ${\bf m^{L}}, {\bf m^{D}}, {\bf m^{R}}$ are Lorentz and $SU(2)_L\times U(1)$ singlet $3\times 3$ complex matrices. ${\bf m^{D}}$ preserves lepton number, but violates $SU(2)_L \times U(1)$ since it connects an $SU(2)_L$ doublet with a singlet and thus can only be generated after spontaneous symmetry breaking. This lepton number conserving mass is often referred to as a Dirac mass. Both ${\bf m^{L}}$ and ${\bf m^{R}}$ violate the individual lepton numbers by two units and are referred to as Majorana masses. The presence of ${\bf m^{L}}$ also requires $SU(2)_L\times U(1)$ symmetry breaking since it connects left handed neutrinos which are parts of an $SU(2)_L$ doublet. In the Standard Model its generation requires a mass dimension five operator. On the other hand, since ${\bf m^{R}}$ connects $SU(2)_L\times U(1)$ singlet right handed neutrino fields, it can appear even in the absence of $SU(2)_L\times U(1)$ symmetry breaking. 

The matrix $C$ appearing in the Majorana neutrino mass terms is there to preserve Lorentz invariance. This matrix satisfies
\be
C C^\dagger =1 ~~;~~C=C^T ~~;~~ C\gamma^*_\mu C^{-1}=-\gamma_\mu ~.
\ee
The specific form for $C$ is representation dependent. In the Dirac representation, $C=i\gamma_2\gamma^0$. The charge conjugate field $\psi^c$ is defined as 
\be
\psi^c(x)= C\overline{\psi}^T(x)~~;~~\overline{\psi^c}=\psi^T C ~.
\ee

Using the charge conjugated fields, one can write 
\be
\overline{\nu^{(0)}_{Ri}}\nu^{(0)}_{La}=\frac{1}{2}[\overline{\nu^{(0)}_{Ri}} \nu^{(0)}_{La}+\overline{(\nu_{La}^{(0)})^c}(\nu_{Ri}^{(0)})^c]
\ee
and thus it follows that the neutrino mass term can be cast in the compact form as
\be
{\cal L}^\nu_{mass}=-\frac{1}{2}\left(\overline{(\nu_{L}^{(0)})^c} ~\overline{\nu^{(0)}_R}\right) \pmatrix{{\bf m^L}&({\bf m^D})^T\cr
{\bf m^D}&{\bf m^R}}\left(\begin{array}{r}\nu_L^{(0)}\\
(\nu_R^{(0)})^c\end{array}\right) + h.c. ~.
\ee
For 3 generations of neutrinos, the six mass eigenvalues, $m_{\nu A}$, are obtained from the diagonalization of the $6\times 6$ matrix
\be
{\bf {M^\nu} } = \pmatrix{{\bf m^L}&({\bf m^{D}})^T\cr
{\bf m^D}&{\bf m^R}} ~.
\ee
Since ${\bf {M^\nu}}$ is symmetric, it can be diagonalized using a unitary transformation
\be
{\bf M^{\nu}_{diag}}=(\bf U)^\dagger {\bf M^\nu} {\bf U} ~,
\ee
where ${\bf U}$ is a $6\times 6$ unitary matrix. This diagonalization is implemented via the basis change on the original neutrino fields organized as the 6 dimensional column vector 
\be
N^{(0)}_L= \pmatrix{\nu_L^{(0)}\cr
(\nu_R^{(0)})^c}~~;~~N_R^{(0)}= \pmatrix{(\nu_L^{(0)})^c\cr
\nu_R^{(0)}}
\ee
to the new neutrino fields (without superscripts) defined as
\be
N^{(0)}_L={\bf U} N_L~~;~~N^{(0)}_R ={\bf U} N_R ~,
\ee
where 
\be
N_L= \pmatrix{\nu_L\cr
(\nu_R)^c}~~;~~N_R=\pmatrix{(\nu_L)^c\cr
\nu_R} ~.
\ee

The neutrino mass term takes the form
\be
{\cal L}_{mass}^\nu = -\frac{1}{2} \sum_{A=1}^6 m_{\nu A}[\nu_A^T C\nu_A + \overline{\nu_A}C\overline{\nu_A^T}]=-\sum_{A=1}^6 m_{\nu A}\overline{\nu_A}\nu_A ~,
\ee
where $m_{\nu A}$ are the Majorana neutrino masses. Note that these mass diagonal fields are Majorana (self-conjugate) fields satisfying $\nu_A = (N_L + N_R )_A = \nu_A^c ~~;~~ A=1,...,6$.

As a consequence of the diagonalization of the neutrino mass matrix, the charged current when written in terms of the mass diagonal fields takes the form
\be
J_{-\mu }^{lept}=(J_{+\mu }^{lept})^\dagger =\sum_{a=1}^3 \sum_{A=1}^6\overline{\ell_{La}}\gamma_\mu V_{aA}\nu_{LA} ~,
\ee
where
\be
V_{aA}=\sum_{b=1}^3 (A^{-1}_L)_{ab} U_{bA} ~.
\ee
Thus the neutrino fields appearing in the charged weak current are given by
\be
\nu^{(0)}_{La}=\sum_{A=1}^6 U_{ aA}\nu_{LA} ~,
\ee
where $\nu_{LA}$ is the $A^{th}$ component of $N_L$.  Note that since both ${\bf A_L}$ and ${\bf U}$ are unitary matrices, it follows that 
\be
\sum_{A=1}^6 V_{aA}V_{bA}^* =\delta_{ab} ~.
\label{uc}
\ee

Since a nonzero Majorana mass matrix ${\bf m^R}$ does not require $SU(2)_L\times U(1)$ symmetry breaking, it is naturally characterized by a much larger scale, $M_R$, than the matrices ${\bf m^D}$ and ${\bf m^L}$ whose nontrivial values do require $SU(2)_L\times U(1)$ symmetry breaking. In fact, since the presence of ${\bf m^L}$ requires a mass dimension five operator, it is naturally much smaller than both $M_R$ and the elements of ${\bf m^D}$ which are expected to be somewhere of the order of the charged lepton mass to the $W$ mass. Thus for simplicity, one can set ${\bf m^L}$ to zero, while taking the elements of ${\bf m^D}$, characterized by a scale $m_D$, to be much less than $M_R$, the scale of the elements of ${\bf m^R}$. So doing, one finds on diagonalization of the $6\times 6$ neutrino mass matrix that 3 of the eigenvalues are crudely given by 
\be
m_{\nu a} \sim \frac{m_D^2}{M_R}<<m_D~~;~~a=1,2,3  ~,
\ee
while the other 3 eigenvalues are roughly 
\be
m_{\nu i} \sim M_R ~~;~~i=4,5,6 ~.
\ee
This constitutes the so called see-saw mechanism\cite{Y} and provides a natural explanation of the smallness of the 3 light neutrino masses. Moreover, the elements of the mixing matrix are characterized by the $M_R$ mass dependence 
\bea
U_{ab}&\sim& {\cal O}(1)~~;~~a,b =1,2,3 \nonumber\\
U_{ij}&\sim& {\cal O}(1)~~;~~i,j =4,5,6 \nonumber\\
U_{ia}&\sim& U^L_{ai}\sim {\cal O}(\frac{m_D}{M_R})~~;~~ a=1,2,3~;~i=4,5,6 ~.
\eea
Since the charged lepton mixing matrix is independent of $M_R$, one finds that elements of the mixing matrix appearing in the charged current has the $M_R$ mass dependence 
\bea
V_{ab}&\sim& {\cal O}(1)~~;~~a,b =1,2,3 \nonumber\\
V_{ai}&\sim& {\cal O}(\frac{m_D}{M_R})~~;~~ a=1,2,3~;~i=4,5,6 ~.
\eea

Note that the states $\nu_a~~;~~a=1,2,3$, are predominately composed, up to corrections of order $\frac{m_D}{M_R}<<1$, of the neutrino fields $\nu^{(0)}_{La}$ which are the ones participating in the weak interaction. Thus for most practical applications, one can simply neglect the heavy neutrino fields which are primarily composed of the fields $\nu^{(0)}_{Ri}~~;~~i=4,5,6$, which are electroweak singlets. As we shall see, however, when considering the case of muonium-antimuonium oscillations, the heavy neutrino states will yield contributions of the same order as the light neutrino states and their presence cannot be neglected. 

The evaluation of the graphs of Fig. 1 contributing to the muonium-antimuonium mixing is straightforward. Taking the external leg momenta to vanish, the result of the calculation can be encapsulated in an effective Lagrangian\cite{FW} of the form
\be 
{\cal L}_{eff} =\frac{G_{\overline{M} M}}{\sqrt{2}} [\overline\mu \gamma^\nu (1-\gamma_5) e][\overline\mu \gamma_\nu (1-\gamma_5) e] ~.
\label{FW1}
\ee
An explicit unitary gauge calculation yields 
\be
\frac{G_{\overline{M}M}}{\sqrt{2}}=-\frac{G_F^2 M_W^2}{16\pi^2}\left[\sum_{A =1}^6 (V_{\mu A}V^\dagger_{e A})^2 S(x_A)+\sum_{A,B=1; A\ne B}^6(V_{\mu A}V^\dagger_{e A})(V_{\mu B}V^\dagger_{e B})T(x_A, x_B)\right]
\label{G} ~,
\ee
where $G_F \simeq 1.16 \times 10^{-5} ~{\rm GeV^{-2}}$ 
is the Fermi scale and $x_A =\frac{m_{\nu A}^2}{M_W^2}~~;~~A=1,...,6$. Here
\be
S(x)=\frac{x^3 -11x^2 +4x}{4(1-x)^2}-\frac{3x^3}{2(1-x)^3}\ell n (x)
\ee
is the Inami-Lin\cite{IL} function and 
\be
T(x_A,x_B) =x_Ax_B \left( \frac{J(x_A)-J(x_B)}{x_A-x_B} \right) =T(x_B,x_A)
\ee
with
\be
J(x)= \frac{ (x^2-8x+4)}{4(1-x)^2}\ell n (x)-\frac{3}{4}\frac{1}{(1-x)} ~.
\ee
In obtaining this result judicious use was made of the unitarity of the charged current mixing matrix elements (c.f. Eq.(\ref{uc})). As a consequence of the fermi statistics, the graphs of figures (a) and (b) gave identical contributions to the effective Lagrangian, while the graphs of figures (c) and (d) gave canceling contributions. Note that in order for the neutrinos in the intermediate state to give a nonvanishing contribution to the effective Lagrangian, they must be massive as well as exhibit a nontrivial mixing with both the electron and the muon in the charged current. 

Muonium (antimuonium) is a nonrelativistic Coulombic bound state of an electron and an anti-muon (positron and muon). The nontrivial mixing between the muonium ($|M>$ ) and antimuonium ($|\overline{M}>$) states is encapsulated in the effective Lagrangian of Eq. (\ref{FW1}) and leads to the mass diagonal states given by the linear combinations
\be
|M_\pm>=\frac{1}{\sqrt{2(1+|\epsilon|^2)}}[(1+\epsilon)|M> \pm (1-\epsilon)|\overline{M}>]
\ee
where 
\be
\epsilon = \frac{\sqrt{{\cal M}_{M\overline{M}}}-\sqrt{{\cal M}_{\overline{M}M}}}{\sqrt{{\cal M}_{M\overline{M}}}+\sqrt{{\cal M}_{\overline{M}M}}}
\ee
with
\be
{\cal M}_{\overline{M}M} = \frac{<\overline{M}|-\int d^3 r {\cal L}_{\rm eff}|M>}{\sqrt{<M|M><\overline{M}|\overline{M}>}} ~~;~~
{\cal M}_{M\overline{M}} = \frac{<M|-\int d^3 r {\cal L}_{\rm eff}|\overline{M}>}{\sqrt{<M|M><\overline{M}|\overline{M}>}} ~. 
\ee
Since the neutrino sector is expected to be CP violating, these will be independent,  complex matrix elements. If the neutrino sector conserves CP, with $|M>$ and $|\overline{M}>$ CP conjugate states, then ${\cal M}_{M\overline{M}}={\cal M}_{\overline{M}M}$ and $\epsilon =0$. In general, the magnitude of the mass splitting between the two mass eigenstates is 
\be
|\Delta M | = 2 |{\rm Re}\sqrt{{\cal M}_{\overline{M}M}{\cal M}_{M\overline{M}}}| ~.
\ee
Since muonium and antimuonium are linear combinations of the mass diagonal states, an initially prepared muonium or antimuonium state will undergo oscillations into one another as a function of time. The muonium-antimuonium oscillation time scale, $\tau_{\overline{M}M}$, is given by
\be
\frac{1}{\tau_{\overline{M}M}}=| \Delta M |  ~.
\ee

A nonrelativistic reduction of the effective Lagrangian of Eq. (\ref{FW1}) produces the local, complex effective potential
\be
V_{\rm eff}(\vec{r})= 8\frac{G_{\overline{M}M}}{\sqrt{2}} \delta^3(\vec{r}) ~.
\ee
Taking the muonium (antimuonium) to be in their respective Coulombic ground states, $\phi_{100}(\vec{r})=\frac{1}{\sqrt{\pi a_0^3}}e^{-r/a_{\overline{M}M}}$, where $a_{\overline{M}M} =\frac{1}{m_{\rm red} \alpha}$ is the muonium Bohr radius with $m_{\rm red}=\frac{m_e m_\mu}{m_e +m_\mu}\simeq m_e$ the reduced mass of muonium, it follows that
\be
\frac{1}{\tau_{\overline{M}M}}\simeq 2\int d^3 r \phi^*_{100}(\vec{r})|{\rm Re}~V_{\rm eff}(\vec{r})|\phi_{100}(\vec{r})
= 16\frac{|{\rm Re}~G_{\overline{M}M}|}{\sqrt{2}}|\phi_{100}(0)|^2 
= \frac{16}{\pi}\frac{|{\rm Re}~G_{\overline{M}M}|}{\sqrt{2}}\frac{1}{a_{\overline{M}M}^3} ~.
\ee
Thus we secure an oscillation time scale  
\be
\frac{1}{\tau_{\overline{M}M}}\simeq \frac{16}{\pi}\frac{|{\rm Re}~G_{\overline{M}M}|}{\sqrt{2}}m_e^3 \alpha^3 ~,
\ee
with $\frac{ G_{\overline{M}M}}{\sqrt{2}}$ given by Eq. (\ref{G}).

The present experimental limit\cite{W} on the non-observation of muonium-antimuonium oscillations translates into the bound  $|{\rm Re}~G_{\overline{M}M}| \leq 3.0 \times 10^{-5}G_F$. This limit can then be used to construct a crude lower bound on $M_R$. For the case when the neutrino masses arise from a seesaw mechanism and taking $m_D$ to be of order $M_W$, the $M_R$ dependence of $G_{\overline{M}M}$ is obtained from Eq.(\ref{G}) as:
\bea
|{\rm Re}~G_{\overline{M}M}|&\sim & \frac{G_F^2M_W^4}{M_R^2}~~;~~ A=a=1,2,3~;~ B=b=1,2,3 \nonumber \\
|{\rm Re}~G_{\overline{M}M}|&\sim & \frac{G_F^2M_W^4}{M_R^2}~~;~~ A=i=4,5,6~;~ B=j=4,5,6 \nonumber \\
|{\rm Re}~G_{\overline{M}M}|&\sim & \frac{G_F^2M_W^6}{M_R^4}\ell n(\frac{M_R}{M_W})~~;~~ A=a=1,2,3~;~ B=i=4,5,6 ~.
\eea
That is, the contributions to $|{\rm Re}~G_{\overline{M}M}|$ are of comparable magnitude when the neutrinos propagating in the loop are both either very light $m_a\sim {\cal O}(\frac{m_D^2}{M_R})$ or  very heavy $m_i \sim M_R$, but are suppressed by an additional factor of $\frac{M_W^2}{M_R^2}\ell n (\frac{M_R}{M_W})$ when one of the neutrinos is light and the other neutrino is heavy. For the two light neutrino intermediate state, there is one factor of $\frac{1}{M_R}$ coming from each of the two light masses, while the mixing matrix elements are of order unity. On the other hand, for the case when there are two heavy neutrinos in the intermediate state, there is a factor of $\frac{1}{M_R}$ arising from each of the four mixing matrix elements and two factors of the heavy mass, $M_R$, from the two masses once again resulting in a net factor of $\frac{1}{M_R^2}$. Thus, even though the heavy Majorana neutrinos constitute but  a very small amount (${\cal O}(\frac{m_D}{M_R})$) of the neutrinos which participate in the weak interactions, they still contribute at the same level (in terms of $M_R$ dependence) as the light neutrinos which are the principle components of the weak interaction neutrino fields.

Unfortunately, the current experimental bound on $|{\rm Re}~G_{\overline{M}M}|$ is not particularly restrictive yielding the modest bound
\be
M_R > {\cal O}(10^4 ~\rm{GeV}) ~.
\ee
We stress that this bound is at best an order of magnitude estimate since we are retaining only the mass dependence on $M_R$ and neglecting all numerical dependence on the mixing angles and CP violating phases in $V_{aA}$.

\vspace*{.3in}
\noindent
This work was supported in part by the U.S. Department 
of Energy under grant DE-FG02-91ER40681 (Task B).

\end{document}